# The transition to classical chaos in a coupled quantum system through continuous measurement

Shohini Ghose,* Paul Alsing, and Ivan Deutsch

*Department of Physics and Astronomy, University of New Mexico, Albuquerque, New Mexico 87131*

Tanmoy Bhattacharya, Salman Habib, and Kurt Jacobs†

*T-8 Theoretical Division, MS B285, Los Alamos National Laboratory, Los Alamos, New Mexico 87545*

(Dated: September 10, 2003)

Continuous observation of a quantum system yields a measurement record that faithfully reproduces the classically predicted trajectory provided that the measurement is sufficiently strong to localize the state in phase space but weak enough that quantum backaction noise is negligible. We investigate the conditions under which classical dynamics emerges, via continuous position measurement, for a particle moving in a harmonic well with its position coupled to internal spin. As a consequence of this coupling we find that classical dynamics emerges only when the position and spin actions are both large compared to $\hbar$. These conditions are quantified by placing bounds on the size of the covariance matrix which describes the delocalized quantum coherence over extended regions of phase space. From this result it follows that a mixed quantum-classical regime (where one subsystem can be treated classically and the other not) does not exist for a continuously observed spin 1/2 particle. When the conditions for classicallity are satisfied (in the large-spin limit), the quantum trajectories reproduce both the classical periodic orbits as well as the classically chaotic phase space regions. As a quantitative test of this convergence we compute the largest Lyapunov exponent directly from the measured quantum trajectories and show that it agrees with the classical value.

## I. INTRODUCTION

The distinct dynamical predictions of quantum and classical mechanics for a given Hamiltonian have been well known since the inception of quantum theory. Although one might naively expect "macroscopic systems" (i.e. ones whose characteristic actions, $I$, are large compared to $\hbar$) to behave classically, for systems with Hamiltonian chaos, Berry and Balazs have argued that the semiclassical approximation may break down in an exceedingly short time, logarithmic in $I/\hbar$ [1]. Understanding how classical chaotic behavior emerges from the underlying quantum description is a fundamental problem in physics. The study of quantum nonlinear dynamics for application to quantum information processing [2] and feedback control [3] provides further motivation to pursue a deeper understanding of the quantum to classical transition in chaotic systems. Finally, the experimental state of the art is rapidly progressing to the situation where individual quantum systems can be monitored in a controlled way [4, 5, 6] necessitating a parallel theoretical development.

In previous studies, the quantum to classical transition was analyzed by comparing distributions in phase space [7]. It has been shown that the process of decoherence due to interaction with an environment can suppress quantum interference so that the quantum quasiprobability distribution remains close to the corresponding classical phase space distribution [8]. Here we extend a different approach taken in Ref. [9] to define the emergence of classical dynamics. Given an experiment in which the dynamical system is continuously observed, we ask under what conditions the measurement record is faithfully predicted by the classical dynamical equations of motion (e.g. Hamilton's equations). In a fundamentally quantum system the continuous measurement has two basic effects: (i) through knowledge gained in the observation the state is localized to within the resolution of the measurement, (ii) backaction noise is imparted to the conjugate variables consistent with the quantum information-disturbance relations. Classical dynamics will provide a good approximation to the measurement record only when the localization is sufficiently strong so that trajectories can be defined in phase space and the backaction noise is sufficiently weak so that these trajectories are barely disturbed. In general, this balance can be struck for a sufficiently macroscopic system. The scale of action relative to $\hbar$ at which this occurs characterizes the quantum/classical dynamical boundary.

A quantitative description of the time evolving continuous measurement record can be made using the quantum trajectory formalism [10]. Bhattacharya, Habib, and Jacobs [9] were able to find the conditions that achieve the strong-localization/weak-backaction balance. The system they studied was the Duffing oscillator — a driven nonlinear system with one dynamical degree of freedom. We seek to generalize this analysis to coupled systems with multiple degrees of freedom and no external classical driving force. New questions arise for such coupled quantum systems. Since the scale of the action relative to $\hbar$ can differ for the different subsystems, one can explore a regime where one degree of freedom has a large action while another is deeply quantum. It is known that such approximate mixed quantum/classical, or "semiclassical", systems can exhibit signatures of chaos [11, 12, 13, 14]: We wish to investigate whether such a description can apply in a real experiment. Another interesting question is whether the quantum entanglement of the different degrees of freedom plays a role in the approach to the classical regime.

---

*Electronic address: sghose1@unm.edu
†Present Address: School of Science, Griffith University, Nathan 4111, Australia

We consider a dynamical system consisting of a particle whose motion in a harmonic well is coupled to its spin. This has wide applicability to a variety of phenomena including generalizations of the Jaynes-Cummings model in quantum optics (coupling of an atomic pseudo-spin to a harmonic mode of the electromagentic field) [15], the spin-boson model of condensed matter systems (e.g. polariton transport) [16, 17], and the motion of ultracold atoms in magneto-optical traps [18]. In previous work [19], we considered an integrable regime in which the Hamiltonian exhibits only regular motion. In the current paper we extend our analysis to the classically chaotic regime. In particular, we study when the classical Lyapunov exponents are recovered from the quantum trajectories. This gives an unambiguous signature of the emergent chaotic behavior.

The paper is organized as follows. We start by describing the coupled motion-spin Hamiltonian in Sec. II. The evolution of the system conditioned on a weak continuous measurement can be described using a stochastic Schrödinger equation as outlined in Sec. III A. We present our numerical results for the evolution of the measured quantum system, starting with the spin-1/2 system (Sec. III B), and then moving to the large spin limit (Sec. III C). Analytical conditions for the recovery of classical dynamics are obtained in Sec. IV by bounding the nonclassical covariance matrix and thereby showing that corrections to the classical trajectories always remain small. In Sec. V we compute the largest Lyapunov exponent of the quantum trajectories and compare it to the classical value in order to quantitatively demonstrate the emergence of classical chaos. We conclude with a brief summary of our primary results in Sec. VI.

## II. THE COUPLED SYSTEM OF SPIN AND MOTION

The Hamiltonian we consider here is

$$\hat{H} = \frac{\hat{p}^2}{2m} + \frac{1}{2}m\omega^2 \hat{z}^2 + b\hat{z}\hat{J}_z + c\hat{J}_x, \qquad (1)$$

where $\hat{z}$ is the position operator of a particle of mass $m$ trapped in a harmonic well of frequency $\omega$ and $\hat{J}_i$ are the components of the particle's spin angular momentum. In addition to the trap, the spin is coupled to an effective magnetic field with a constant transverse ($x$-direction) component and gradient along the longitudinal ($z$-) direction. We can make the analogy to a classical Hamiltonian by replacing the spin with a classical magnetic moment of magnitude $\boldsymbol{\mu} = \gamma \mathbf{J}$, where $\gamma$ is the gyromagnetic ratio, interacting with the local field $\mathbf{B}(z) = -(c\mathbf{e}_x + bz\mathbf{e}_z)/\gamma$ via $-\boldsymbol{\mu} \cdot \mathbf{B}(z)$. For this classical analog system, coupling between the direction of the magnetic moment and the position of the particle in the wells can lead to chaotic motion in a spatially inhomogeneous field [20, 21].

The expectation values of the Heisenberg equations obtained from Eq. (1) are

$$\begin{aligned}
\frac{d\langle \hat{z} \rangle}{dt} &= \frac{\langle \hat{p} \rangle}{m}, \\
\frac{d\langle \hat{p} \rangle}{dt} &= -m\omega^2 \langle \hat{z} \rangle - b\langle \hat{J}_z \rangle, \\
\frac{d\langle \hat{\mathbf{J}} \rangle}{dt} &= \gamma \langle \hat{\mathbf{J}} \rangle \times \mathbf{B}(\langle \hat{z} \rangle) + \gamma \mathbf{C}_{\hat{\mathbf{J}} \times \mathbf{B}(\hat{z})},
\end{aligned} \qquad (2)$$

where

$$C_{\hat{\mathbf{J}} \times \mathbf{B}(\hat{z})} = \langle \hat{\mathbf{J}} \times \mathbf{B}(\hat{z}) \rangle - \langle \hat{\mathbf{J}} \rangle \times \mathbf{B}(\langle \hat{z} \rangle) \qquad (3)$$

is the covariance or the second cumulant. In general, these correlations are non-zero, so that the quantum expectation values do not follow the classical trajectories. As we will see, in the small $\hbar$ limit, continuous measurement can act to damp the higher order cumulants with negligible quantum backaction noise thereby recovering classical dynamics.

A special case to consider is when the action associated with center-of-mass dynamics is large enough such that, were there no coupling between the two degrees of freedom, the motion in the harmonic wells could be treated classically while the uncoupled spin would still be deeply quantum. In the coupled system, should we continue to assume that the motional subsystem can be treated classically and treat the position and momentum operators approximately as c-numbers, then $C_{\mathbf{J} \times \mathbf{B}} \approx 0$. This leads to the "semiclassical" Heisenberg equations of motion, which have *exactly* the same form as the classical Hamilton's equations with $\langle \hat{z} \rangle \to z$, etc. If this approximation were correct it would imply that dynamics in this regime may also exhibit chaos as has been studied in various contexts [11, 12, 13, 14]. The validity of this approximation and the resulting chaos has been questioned in [22, 23]. One of our goals in this article is to investigate whether this "semiclassical chaos" can be recovered in the quantum trajectories, obtained when the system is weakly observed.

## III. CONTINUOUS MEASUREMENT OF POSITION

### A. Conditioned Dynamics

Using the formalism of generalized measurements, we model a weak continuous observation of the particle's position via a stochastic Schrödinger equation (SSE) that describes the evolution of the unnormalized wave function $\left|\tilde{\psi}\right\rangle$, conditioned on a record of the position,

$$d\left|\tilde{\psi}\right\rangle = \left\{\left(\frac{1}{i\hbar}H - k\ z^2\right)dt + \left(4k\langle z\rangle dt + \sqrt{2k}dW\right)z\right\}\left|\tilde{\psi}\right\rangle. \quad (4)$$

This general form of the SSE for a continuous measurement, described by a Wiener process $dW$ of "strength k" and yielding a record $\langle z\rangle + (8k)^{-1/2}\ dW/dt$ has been previously derived for the specific case in which the position of a moving mirror is monitored by an optical probe [24, 25]. Scott and Milburn [26] obtained a similar equation for simultaneous measurements of position and momentum using previous results on continuous position measurement operators [27, 28].

We evolve the SSE numerically using a "split operator" method [29, 30]

$$|\psi(t+dt)\rangle = e^{\frac{-Mdt}{2}}e^{-\frac{i}{\hbar}Hdt}e^{\frac{-Mdt}{2}}|\psi(t)\rangle, \quad (5)$$

where $|\psi\rangle$ and $\left|\tilde{\psi}\right\rangle$ differ only in normalization. The exponentiation of the Hamiltonian is written in Cayley form [30]

$$e^{-\frac{i}{\hbar}Hdt} \simeq \left(\frac{1-\frac{i}{4\hbar}Tdt}{1+\frac{i}{4\hbar}Tdt}\right)\left(\frac{1-\frac{i}{2\hbar}Vdt}{1+\frac{i}{2\hbar}Vdt}\right)\left(\frac{1-\frac{i}{4\hbar}Tdt}{1+\frac{i}{4\hbar}Tdt}\right), \quad (6)$$

where $T$ is the kinetic energy, $V$ the potential, and $M = k[z - (\langle z\rangle + dW/\sqrt{8k}dt)]^2$ represents the conditioning and backaction due to coupling of the particle to the measurement apparatus. The potential operator, block tri-diagonal in the basis of position and $J_z$ eigenstates, can be calculated using efficient algorithms for inverting such matrices [31, 32]. As usual, the kinetic term is applied in the momentum basis using Fast Fourier Transforms. In order to increase the efficiency of our numerical code we use a small grid in position and momentum adaptively centered around the location of the wave function.

### B. The spin-1/2 system

We start by investigating the conditioned evolution of a spin $J = 1/2$ system. We choose the initial state to be a product of a coherent state for the motion (position and momentum phase plane) and a spin coherent state (direction $(\theta, \phi)$ on the Bloch sphere),

$$|\psi(0)\rangle = |\alpha = z + ip\rangle|\theta, \phi\rangle. \quad (7)$$

We pick the spin direction to be along $x$ so that $(\theta, \phi) = (\pi/2, 0)$ (though any other direction would have been equally suitable), the initial momentum to be zero, and the initial position to be $z(0) \approx 38z_g$, with $b = m\omega^2\Delta z/J$, and $\Delta z \approx 22z_g$ where $z_g = \sqrt{\hbar/2m\omega}$ is the width of the harmonic oscillator ground state. For these choices, the action in the motional phase space $I_0 = m\omega\Delta z^2 = 250\hbar$. This puts us in the mixed quantum-classical regime described in Sec. II. The transverse magnetic field is chosen so that $c = 200E_g/J$, where $E_g = \hbar\omega/2$ is the ground state energy. We pick a measurement strength $k = \omega/20z_g^2$ that satisfies the inequalities for strong-localization/weak-backaction found in Ref. [9] in the absence of coupling to the spin. This enables us to study how the coupling to the spin changes the effect of the measurement.

Differences between quantum and classical trajectories arise from two possible sources, nonclassical initial states and nonclassical dynamics. For the integrable regime that we studied previously [19], the system was linear, and therefore the quantum and classical propagators were *equivalent*. Thus, only the difference between the quantum and classical initial *state* was responsible for any disparities between the quantum and classical trajectories [19]. For the nonlinear dynamics considered here, the classical and quantum equations of motion for the higher cumulants differ, and because of severe quantization effects for small spin, the wave function distribution is far from gaussian, making the cumulant expansion of limited utility.

The time evolving measurement record of the position of the spin-1/2 system is compared to the trajectory predicted by the classical equations of motion in Fig. 1. Consider first the classical dynamics. Note that even though there is a transverse magnetic field, the motion is regular, not chaotic. This can be understood by writing the classical equations of motion as,

$$\begin{aligned}
\frac{d\tilde{z}}{d\tau} &= \tilde{p}, \\
\frac{d\tilde{p}}{d\tau} &= -\tilde{z} - n_z, \\
\frac{dn_x}{d\tau} &= -\frac{I_0}{J}\tilde{z}n_y, \\
\frac{dn_y}{d\tau} &= \frac{I_0}{J}\tilde{z}n_x - \tilde{c}\frac{I_0}{J}n_z, \\
\frac{dn_z}{d\tau} &= \tilde{c}\frac{I_0}{J}n_y,
\end{aligned} \quad (8)$$

where

$$\begin{aligned}
\tilde{z} &= z/\Delta z, \\
\tilde{p} &= p/m\omega\Delta z, \\
I_0 &= m\omega\Delta z^2, \\
\tilde{c} &= cJ/m\omega^2\Delta z^2, \\
\tau &= \omega t,
\end{aligned} \quad (9)$$

where **n** is the direction of the magnetic moment. We can see from these equations that the dynamics depends on the ratio of the actions of the coupled subsystems. In the classical limit, both external and internal actions $I_0$ and $J$ are infinitely large relative to $\hbar$, but the ratio of the two is finite. For the "semiclassical" regime considered here, $I_0/J = 500$. The large disparity between the actions of the external and internal dynamics separates the time scales of the position and magnetic moment evolutions, effectively decoupling the spin from its motion in the well. This leads to a regime where the magnetic moment can adiabatically follow the changing magnetic field direction. The angle between the magnetic moment direction and the local magnetic field becomes an additional constant of motion apart from the energy, giving rise to integrable motion.

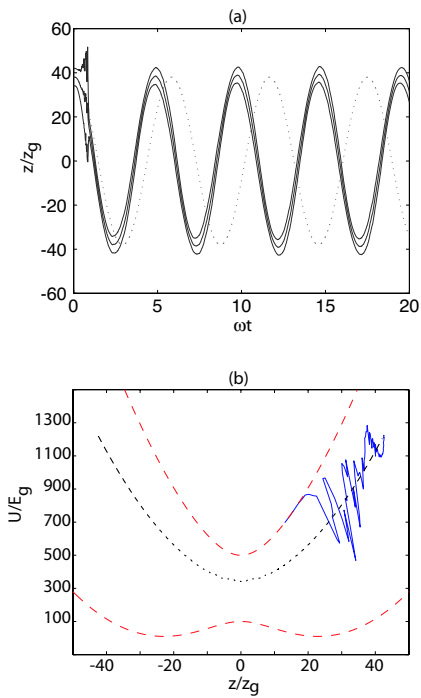

FIG. 1: (a) Mean position of the measured spin-1/2 system (solid) in a single quantum trajectory with $\Delta z \approx 22 z_g$, $c/J = 200 E_g$, $k = \omega/20 z_g^2$. Outer solid curves show the variance of the wave function. The measurement backaction causes the quantum trajectory to diverge from the classical (dotted, black) trajectory. This is because part of the wave function moves along the upper adiabatic potential while the rest moves along the lower adiabatic potential (dashed red curves in (b)). Eventually, the measurement collapses the wave function into the upper or lower potential (solid blue line in (b)). The classical motion is along the dashed-dotted potential in (b).

Fig. 1 also shows a typical measured quantum trajectory of the spin-1/2 system. After a very short time, it fails to follow the classical adiabatic motion described above. This behavior can be understood in a manner similar to that presented in our previous study of linear dynamics, with no transverse magnetic field [19]. Here, the initial wave function can be decomposed in the basis of adiabatic eigenstates $|\pm(z)\rangle$, obtained by diagonalizing the total potential at each position,

$$|\psi(0)\rangle = |\alpha\rangle |\theta, \phi\rangle = (\phi_+(z)|+(z)\rangle + \phi_-(z)|-(z)\rangle). \quad (10)$$

At the initial time, the $\phi_+(z)$ and $\phi_-(z)$ components overlap in space, but are pulled apart by the differential force of the upper and lower adiabatic potentials (dashed lines in Fig. 1(a)). As the wavepacket splits, the overlap between $\phi_+$ and $\phi_-$ gradually decreases and eventually, when the position measurement can resolve the two spatially separated components, the state is projected into one of the two quantum adiabatic eigenstates. This contrasts with the classical adiabatic motion which moves on an *average* of these two quantum potentials (Fig. 1(b)). Thus, due to the entanglement between motion and spin, the weak measurement of position results in a projective measurement of the spin, much like the situation for the spin-1/2 particle with $c = 0$ [19]. The difference here is that the pointer basis associated with the measurement apparatus is the adiabatic basis rather than the magnetic sublevels associated with the space-fixed quantization axis.

We have thus shown that a continuously observed system in a mixed quantum-classical regime does not follow the classical trajectory. This divergence of the two time series occurs even in the absence of chaos and is simply a statement that there is no smooth classical limit for low-spin systems. To explore the chaotic regime, we must reduce the ratio of the two actions in Eq. (8). However, for a spin-1/2 system, this requires reduction of the external action to values that violate the conditions required for classical dynamics given in Ref. [9]. This means that both the predicted regular and chaotic classical trajectories in the "semiclassical" spin-1/2 description cannot be seen in the measured dynamics.

### C. The large spin limit

The classical equations of motion result in chaotic dynamics when the time scales of the internal and external dynamics are on the same order. With this in mind we set $I_0/J = 5$ and $\tilde{c} = 0.4$ in the equations of motion (Eq. (8)). The classical phase space for these parameters is mixed, with regions of stable motion separated by stochastic layers. We show in this section that this classical chaotic behavior can be recovered from the measured quantum system in the large action limit.

Figs. 2(a,b) show two classical trajectories, one of which is in a regular part of the phase space, and the other in the chaotic region. The corresponding quantum trajectories are shown in Figs. 2(c,d) for a spin with an action $J = 200$ and measurement strength $k = \omega/8 z_g^2$. We pick $\Delta z \approx 45 z_g$ which results in a ratio of characteristic actions in the quantum system that is the same as the classical ratio $I_0/J = 5$. As in the spin-1/2 case, our choice of measurement strength $k$ satisfies the conditions for classicality in [9] had there been no cou-

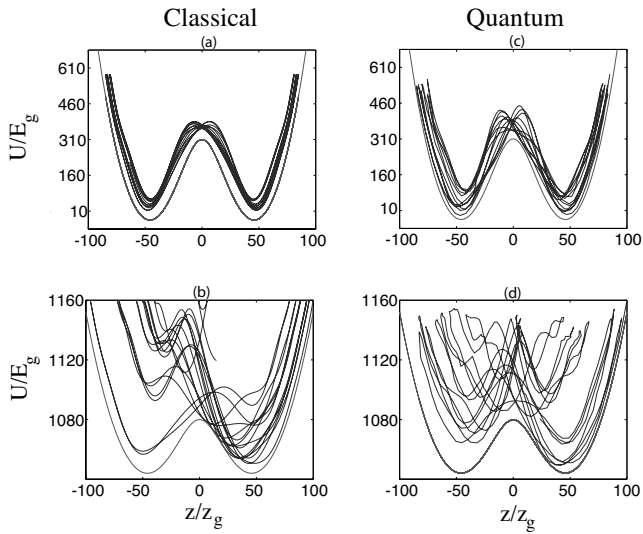

FIG. 2: Regular and chaotic classical trajectories (a,b) are recovered in the quantum trajectories (c,d) with $J = 200\hbar$, $\Delta z \approx 45 z_g$, $k = \omega/8z_g^2$.

pling to the spin. The initial quantum state is chosen to be a product of a coherent state in position and momentum, centered at the classical initial values, and a spin coherent state pointing in the same direction as the initial classical magnetic moment. In Fig. 2, the initial conditions for the regular quantum trajectory were $z(0) = 76z_g, p(0) = 0$ and $(\theta, \phi) = (\pi, 0)$. The initial conditions for the chaotic quantum trajectory were $z(0) = 89z_g, p(0) = 0, (\theta, \phi) = (\pi, 0)$.

The quantum trajectories in Fig. 2 successfully reproduce the classical mixed phase space. This is because for $J = 200$, there are $2J + 1 = 401$ adiabatic potentials rather than just two as in the spin-1/2 case. The initial state is in a superposition of the 401 eigenstates, but has most of its support concentrated on just a few of the adiabatic potentials closest to the local direction of the classical magnetic moment. The differential force is thus very weak and only slightly splits the wavepackets into nonoverlapping components. The position measurement acts only to damp the tails of the distribution where spread is substantial and keeps the wave function localized. It does not, however, strongly project the spin state into a single adiabatic state. A weak measurement of the position also acts as a weak measurement of the spin so that the strong localization and weak backaction conditions can simultaneously be satisfied for both the position and the spin.

A further qualitative example of the quantum trajectories recovering different structures in the mixed classical phase space is shown in Fig. 3. The dots represent a classical surface of section at $E = 0.08E_0 = 0.08m\omega^2\Delta z^2$ with $I_0/J = 2.5$ and $\tilde{c} = 0.4$. The slice is taken at $J_y = 0, dJ_y/dt > 0$. This surface of section shows many islands of regular motion with thin stochastic layers in between. An initial classical trajectory that starts on a regular island cannot cross the stochastic layer that bounds it. In the limit of large actions, the asterisks in Fig. 3 show how quantum trajectories follow different classical periodic orbits. It is possible for the noise due to the measurement to cause a quantum trajectory moving on a periodic orbit to drift into the chaotic region and hence cross a KAM surface. However, as the spin becomes larger and larger, the noise becomes smaller and smaller, eventually becoming negligible in the extreme classical limit and it becomes increasingly unlikely for the quantum trajectory to cross a KAM boundary.

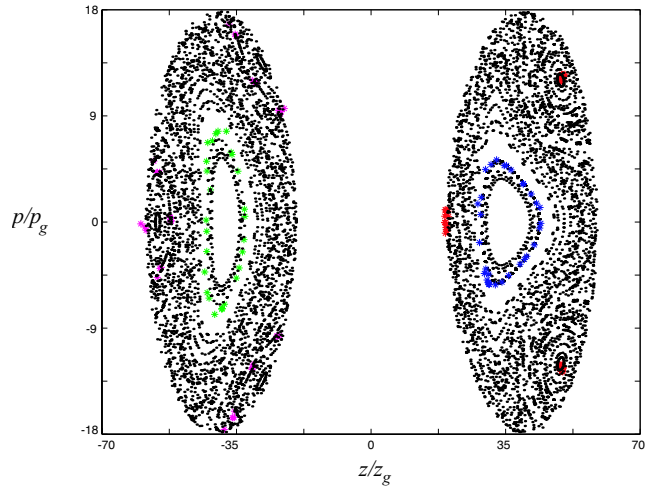

FIG. 3: Quantum trajectories (asterisks) follow the stable islands of the classical surface of section (black dots). Shown are quantum trajectories that reproduce the regular motion around different period-1 fixed points (blue,green), a period-3 orbit (red) and a higher period orbit (magenta).

## IV. CONDITIONS FOR RECOVERING CLASSICAL CHAOS

Classical dynamics is recovered when the mean position, momentum and spin of the measured quantum system follow classical trajectories. The equations of motion for the means of these observables conditioned on the measurement are

$$\begin{aligned}
d\langle \hat{z} \rangle &= \frac{\langle \hat{p} \rangle}{m} dt + \sqrt{8k} C_{zz} dW, \\
d\langle \hat{p} \rangle &= -m\omega^2 \langle \hat{z} \rangle dt - b\langle \hat{J}_z \rangle dt + \sqrt{8k} C_{zp} dW, \\
d\langle \hat{\mathbf{J}} \rangle &= \gamma \langle \hat{\mathbf{J}} \times \mathbf{B}(\hat{z}) \rangle dt + \sqrt{8k} C_{z\mathbf{J}} dW \\
&= \gamma \langle \hat{\mathbf{J}} \rangle \times \mathbf{B}(\langle \hat{z} \rangle) dt + \gamma C_{\hat{\mathbf{J}} \times \mathbf{B}(\hat{z})} dt \\
&\quad + \sqrt{8k} C_{z\mathbf{J}} dW. \quad (11)
\end{aligned}$$

We show how these equations approach the classical equations of motion by imposing the dual conditions of strong localization and weak noise. The measurement must be strong enough to localize the state in phase space so that it resembles a classical point, but weak enough to cause minimal measurement noise or backaction. These dual conditions require the variances and covariances in Eq(11) to remain small relative to the total phase space explored by the motion. Using the approximation that the state remains almost Gaussian in the large action limit, the evolution of the second cumulants can be written in terms of a matrix Ricatti equation,

$$\dot{C}(t) = U + C(t)VC(t) + WC(t) + C(t)W^T, \quad (12)$$

with

$$U = \begin{pmatrix} 0 & 0 & 0 & 0 & 0 \\ 0 & 2\hbar^2 k & 0 & 0 & 0 \\ 0 & 0 & 0 & 0 & 0 \\ 0 & 0 & 0 & 0 & 0 \\ 0 & 0 & 0 & 0 & 0 \end{pmatrix}, \quad V = \begin{pmatrix} -8k & 0 & 0 & 0 & 0 \\ 0 & 0 & 0 & 0 & 0 \\ 0 & 0 & 0 & 0 & 0 \\ 0 & 0 & 0 & 0 & 0 \\ 0 & 0 & 0 & 0 & 0 \end{pmatrix},$$

$$W = \begin{pmatrix} 0 & 1/m & 0 & 0 & 0 \\ -m\omega^2 & 0 & 0 & 0 & -b \\ -b\langle J_y(t)\rangle & 0 & 0 & -b\langle z(t)\rangle & 0 \\ b\langle J_x(t)\rangle & 0 & b\langle z(t)\rangle & 0 & -c \\ 0 & 0 & 0 & c & 0 \end{pmatrix}. \quad (13)$$

where $C$ is the covariance matrix,

$$C = \begin{pmatrix} C_{zz} & C_{zp} & C_{zJ_x} & C_{zJ_y} & C_{zJ_z} \\ C_{zp} & C_{pp} & C_{pJ_x} & C_{pJ_y} & C_{pJ_z} \\ C_{zJ_x} & C_{pJ_x} & C_{J_xJ_x} & C_{J_xJ_y} & C_{J_xJ_z} \\ C_{zJ_y} & C_{pJ_y} & C_{J_xJ_y} & C_{J_yJ_y} & C_{J_yJ_z} \\ C_{zJ_z} & C_{pJ_z} & C_{J_xJ_z} & C_{J_yJ_z} & C_{J_zJ_z} \end{pmatrix}. \quad (14)$$

Unlike the $c = 0$ case [19], we can no longer ignore the $J_x$ and $J_y$ components of the spin. Furthermore, the time-dependence of the matrix $W$ makes it impossible to solve for $C(t)$ analytically.

We can, however, numerically integrate the coupled stochastic equations for the means (Eq. (11)) and the second cumulants (Eq. (12)). We do so by using an explicit Runge-Kutta type algorithm that is strongly convergent to order 1.5 [33]. Fig. 4 shows the solution of the Riccati equation for $C_{zz}$ using the quantum trajectories of Fig. 2. The other cumulants have similar magnitudes. Since the second cumulants remain small relative to the size of the phase space, we expect the solutions of Eq. (11) to agree with the classical solutions at this value of the actions. Our numerical studies showed that this is indeed true (Fig. 2). Furthermore, we have verified that at these large values of the actions, the trajectories obtained by evolving the full SSE agree well with those obtained by solving the equations for the means and second cumulants, indicating that the Gaussian approximation is valid. Hence, numerical solutions of the Riccati equation are a good indication of when the measured dynamics can be approximated classically.

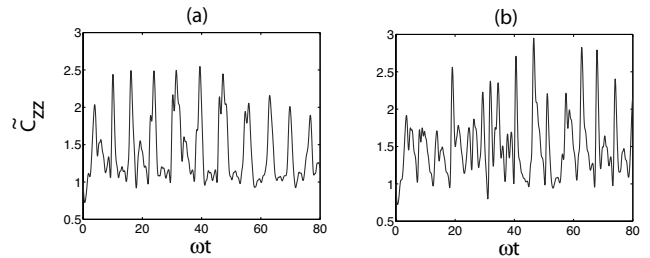

FIG. 4: Solutions of $\tilde{C}_{zz}(t) = C_{zz}(t)/z_g^2$ for the regular and chaotic quantum trajectories of Fig. 2. The maximum cumulant is smaller than the total phase space covered by the motion by a factor of about 100.

## V. QUANTITATIVE RECOVERY OF CLASSICAL CHAOS

Our numerical and analytical studies have shown that qualitative features of the classical trajectories are found in the measurement record of a continuously observed quantum system when the actions are sufficiently large. We would also like to recover some quantitive property of the classical dynamics in order to make a direct comparison. A standard measure of the degree of stochasticity of a classical chaotic system is the largest Lyapunov exponent, describing the average rate of divergence of neighboring trajectories.

We can compute a Lyapunov exponent for the measured quantum trajectories by using a method similar to that employed in classical nonlinear dynamics [34, 35, 36]. In the classical case one chooses a "fiducial trajectory" and calculates the average rate at which this and a neighboring trajectory (nominally an infinitesimal distance away) diverge. In determining the rate numerically, the neighboring trajectory is chosen at a finite, but very small distance, $\epsilon$ from the fiducial. The distance between the fiducial and neighboring trajectory is propagated for a short time $T$ to obtain $d_1(T)$. One then restarts a neighboring trajectory, displaced a distance $\epsilon$ from the fiducial along the direction connecting the fiducial and old neighboring trajectory at time $T$, and propagates the distance again to yield the distance $d_2(T)$. After a long time average, for $N \to \infty$ iterations, the rate of divergence will converge to the largest Lyapunov exponent

$$\lambda_1 = \frac{1}{NT} \sum_{i=1}^{N} \ln \frac{d_i(T)}{\epsilon}. \quad (15)$$

For the quantum trajectories, one can repeat the same procedure, replacing the points that define the classical trajectories with the mean values of the relevant observables. The quantum state, however, is defined by all higher cumulants and these can effect the dynamics. Motivated by the classical analysis, when defining a fiducial and neighboring quantum state, we do it so that they differ only in their means but share exactly the same higher order cumulants. We can achieve this with the help of the phase-space displacement and rotation operators, as described below.

Our numerical procedure for extracting Lyapunov exponents from the continuously observed quantum system is thus as follows. We calculate a fiducial quantum trajectory starting with an initial product coherent state of motion and spin. The neighboring trajectory is chosen by applying the joint displacement-rotation operator to the fiducial at each step $T$. After, say the first iteration, the distance between these quantum trajectories is calculated from the differences in the means,

$$d_1(T) = \sqrt{\delta z(T)^2 + \delta p(T)^2 + \delta \boldsymbol{\mu}(T)^2}. \qquad (16)$$

The neighboring trajectory is then restarted by using the displacement and rotation operators to shift the fiducial state at time $T$ by $\epsilon$ along the direction connecting the the fiducial and original neighboring trajectory at time $T$. For example, the new mean position of the neighboring trajectory, $z'(T)$ is related to that of the fiducial $z(T)$ by

$$\langle z'(T) \rangle = \langle z(T) \rangle + \frac{\delta z(T)}{d_1(T)}\epsilon. \qquad (17)$$

This process is repeated N times, with $N \to \infty$, and the largest Lyapunov exponent is then determined via. Eq. (15).

When the magnitude of the spin $J$ and external action $I_0$ are large (the regime of interest), the Hilbert space dimension of the coupled system grows and the tracking the evolution of the full quantum state becomes numerically intensive. However, we have shown in the previous section that the Gaussian approximation applies in this regime. We can thus use this approximation to efficiently propagate the quantum trajectories and thus compute the Lyapunov exponent for large values of the spin and external action. As a technical aside, when $T$ is very small, the quantum noise due to the measurement can mask the exponential divergence of the quantum trajectories. We can cancel this effect by ensuring that the noise realizations, $dW$, for the fiducial and quantum trajectory are the same.

Fig. 5(a) shows a distribution of the largest classical Lyapunov exponent obtained from 500 fiducial trajectories at an energy of $E = 0.58E_0$ with $I_0/J = 5$ and $\tilde{c} = 0.4$. The Lyapunov exponent, computed using 100 fiducial quantum trajectories with $J = 200\hbar$, $\Delta z \approx 45z_g$ and $k = \omega/8z_g^2$ (Fig. 5(b)), show good agreement with the classical distribution. As discussed above, for numerical efficiency the quantum trajectories were propagated using the coupled equations for the means and second order cumulants. We have verified that for these values of the actions, the trajectories obtained by solving these equations are a good approximation to the exact trajectories obtained by solving the full SSE.

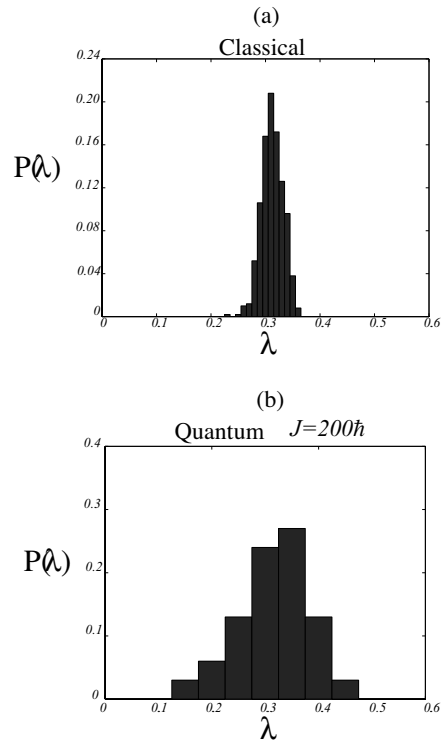

FIG. 5: Distribution of the largest Lyapunov exponent obtained from: (a) classical dynamics using 500 fiducial trajectories at $E = 0.58E_0$ with $I_0/J = 5$ and $\tilde{c} = 0.4$. (b) continuously measured quantum dynamics using 100 fiducial quantum trajectories with $J = 200\hbar$, $\Delta z \approx 45z_g$ and $k = \omega/8z_g^2$. For numerical efficiency, the SSE was integrated by truncating the cumulants at second order.

## VI. SUMMARY

We have studied the conditions under which the measurement record of a continuously observed quantum system can faithfully reproduce the chaotic trajectories predicted by classical mechanics. This represents the first such calculation for the case of *coupled* degrees of freedom – spin and motion – with an undriven Hamiltonian whose classical dynamics can exhibit chaos. In the mixed quantum-classical regime, with large motional action and small spin, the continuous measurement cannot simultaneously satisfy the conditions of strong localization and weak noise, thereby making it impossible to observe "semiclassical chaos". In the large spin limit, both conditions for classicality can be simultaneously satisfied. We computed a Lyapunov exponent directly from the measured quantum trajectories that agrees with the largest classical Lyapunov exponent, thus showing the quantitative correspondence of classical and quantum trajectories. We also obtained general conditions for recovering classical dynamics from the measurement trajectories by studying the evolution of the covariance matrix. These measure the quantum coherence that is delocalized across phase space and thus cause differences between the quan-tum and classical propagators. While we can solve for the covariance matrix analytically in the integrable regime, for the chaotic case we solved the problem numerically.

Whereas coupled degrees on freedom can lead to chaos the classical level, at the quantum level nonseparable Hamiltonians will generally lead to entanglement between the different subsystems (here motion and spin). Entanglement is generally considered to be *the* feature which distinguishes quantum states from their classical counterparts. It is thus natural to explore how the entanglement in our system varies as classical dynamics is recovered in the measured quantum trajectories. As in a pure system entanglement is as good a measure of correlation as the covariance, it might be useful to study the approach to classicality in terms of this complementary variable instead. We plan to explore this possibility in more detail in future work.